\documentstyle[twoside,fleqn,npb,epsfig,axodraw,graphics]{article}
\newcommand{\AmS}{{\protect\the\textfont2
  A\kern-.1667em\lower.5ex\hbox{M}\kern-.125emS}}

\begin{document}
%\begin{frontmatter}

% declarations for front matter
\title{
\rightline{${\normalsize \begin{array}{c} \mathrm{UG-FT-115/00} \\
\mathrm{hep-ph/0011142} 
\\ \mathrm{November\quad 2000}\end{array}}$}  
Top mixing in effective theories  \thanks{Presented at the 5th
Zeuthen Workshop on Elementary Particle 
Theory ``Loops and Legs in Quantum Field Theory'',
Bastei/K\"{o}nigstein, Germany, April 9-14, 2000}}
\author{F. del Aguila and J. Santiago 
\address{
Departamento de F\'{\i}sica Te\'{o}rica
y del Cosmos,  Universidad de Granada.  E-18071 Granada, Spain}}%

\begin{abstract}
We review how top mixing with light quarks constrains new physics
beyond the Standard Model using the effective Lagrangian approach
\end{abstract}

\maketitle

\vspace{-10cm}

\section{Introduction}

The top quark is expected to be the main probe to new physics beyond
the Standard Model (SM) in forthcoming experiments. On one hand its
couplings are not measured with high precision. They are known at the
tens per cent level~\cite{Beneke:2000hk}. 
On the other hand, for example, the Large Hadron
Collider (LHC) at CERN will be a top factory with more than $10^6$
single 
and $10^7$
pair produced top quarks per year. This will stand for a precision in the
determination of the renormalizable $Vtq$ couplings, $V=W,Z$, of
$\sim 1\%$~\cite{Beneke:2000hk}, more than one order of magnitude
better than present accuracy.  
If the new amplitudes scale with the
masses involved in the process, the effective top coupling with the
charm quark would be $\frac{m_t}{m_b}\cdot\frac{m_c}{m_s}\sim 400$
times larger than the bottom coupling to the strange quark, and then a
probe all these times more efficient. Hence it is 
important to know which new physics can show up at the top and how to
discriminate among the different SM extensions.

Our knowledge of the top properties and the precision to be reached in
the near future have been revised in detail  in
Ref.~\cite{Beneke:2000hk} recently. In the following we present the effective
Lagrangian description of top mixing and justify why extra
vector-like quarks can induce its largest
values \cite{apvs:letters}. We then evaluate the new
couplings for the simplest case of an extra quark isosinglet of charge
$\frac{2}{3}$, $T_{L,R}$, and comment on the experimental limits.

\section{Effective Lagrangian descripiton of top mixing}

In order to describe new physics beyond the SM in a model-independent
way we must use an effective Lagrangian \cite{wein}. Assuming
the validity of the SM at the electroweak scale and then that only the
SM fields are light, the effects of any SM gauge extension, including
the possibility of extra dimensions at the TeV scale, are parametrized
by the most general effective Lagrangian involving the SM fields and
preserving the SM symmetries. Imposing the almost exact baryon and
lepton number conservation such a Lagrangian has been written down up
to dimension 6 terms \cite{Buchmuller:1986jz},

\begin{equation}
{\mathcal{L}}_{\mathit{eff}}={\mathcal{L}}_4+
\frac{1}{\Lambda^2}{\mathcal{L}}_6+\ldots.   
\end{equation}
The lowest order part ${\mathcal{L}}_4$ containing all 
renormalizable terms is fixed by the SM. There is no dimension 5
operator allowed by the symmetries; whereas without taking into
account flavour indices there are 81 independent dimension 6 operators
contributing to ${\mathcal{L}}_6$, 40 of them involving quarks of
charge $\frac{2}{3}$.

\begin{table*}[!t]
\caption{Dimension 6 operators contributing to renormalizable $Vtq$
couplings after SSB.}
\label{operators:tab}\vspace{0.5cm}
\begin{tabular}{lll}
${\mathcal{O}}_{\phi q}^{(1)}= 
(\phi^\dagger i
D_\mu\phi)(\bar{q}\gamma^\mu q)$
& ${\mathcal{O}}_{\phi u}= (\phi^\dagger i
D_\mu\phi)(\bar{u}\gamma^\mu u)$&
${\mathcal{O}}_{\phi W}=\frac{1}{2}(\phi^\dagger
\phi)W^I_{\mu\nu}W^{I\mu\nu}  $\\
${\mathcal{O}}_{\phi q}^{(3)}= 
(\phi^\dagger \tau^I i
D_\mu\phi)(\bar{q}\gamma^\mu \tau^I q)$
& ${\mathcal{O}}_{\phi \phi}= (\phi^T \epsilon i
D_\mu\phi)(\bar{u}\gamma^\mu d)$&
${\mathcal{O}}_{\phi B}=\frac{1}{2}(\phi^\dagger
\phi)B_{\mu\nu}B^{\mu\nu} $\\ 
${\mathcal{O}}_{u \phi}= 
(\phi^\dagger \phi)(\bar{q} u \tilde{\phi})$
&
&
${\mathcal{O}}_{W B}=(\phi^\dagger \tau^I
\phi)W^I_{\mu\nu}B^{\mu\nu} $\\ 
${\mathcal{O}}_{d \phi}= 
(\phi^\dagger \phi)(\bar{q} d \phi)$
& &
${\mathcal{O}}_\phi^{(1)}=(\phi^\dagger \phi)  (D_\mu\phi^\dagger
D^\mu\phi)  $\\ 
& &
${\mathcal{O}}_\phi^{(3)}=(\phi^\dagger D^\mu\phi)  (D_\mu\phi^\dagger
\phi) $ 
\end{tabular}
\vspace{0.5cm}
\end{table*}

${\mathcal{L}}_{\mathit{eff}}$ should give a good quantitative
description of the $1\%$ SM deviations, which is the size of the top
mixing $Vtq$ experimentally testable~\cite{Beneke:2000hk}. 
However this Lagrangian compares with
experiment only after electroweak Spontaneous Symmetry Breaking (SSB). Thus
${\mathcal{L}}_6$ gives contributions to dimension 4 operators
${\mathcal{O}}(\frac{v^2}{\Lambda^2})$
, with $v$ the electroweak
vacuum expectation value $\sim 246$ GeV. There are also operators
of dimension 5 ${\mathcal{O}}(\frac{v}{\Lambda^2})$ and the initial
dimension 6 operators
${\mathcal{O}}(\frac{1}{\Lambda^2})$. Experimental strategies must be
designed to disentangle the different contributions. The obvious way is
to start with the simplest vertices. Hence we will concentrate on the
corrected gauge couplings of  dimension 4  describing the top mixing 
with light quarks~\cite{Beneke:2000hk} 

\begin{table*}[t]
% space before first and after last column: 1.5pc
% space between columns: 3.0pc (twice the above)
\setlength{\tabcolsep}{1.5pc}
% -----------------------------------------------------
% adapted from TeX book, p. 241
\newlength{\digitwidth} \settowidth{\digitwidth}{\rm 0}
\catcode`?=\active \def?{\kern\digitwidth}
% -----------------------------------------------------
\caption{Top quark flavour changing branching ratios for
the SM, the two Higgs (2H) model, supersymmetric models (SUSY) without
$\not{\!\mbox{R}}$ and with  R parity breaking and the SM extensions
with  exotic
vector-like quarks~\cite{Beneke:2000hk}. The branching ratio is defined as
$B=\frac{\Gamma} {1.56 \mbox{{\scriptsize GeV}}}$.}
\label{br:models}\vspace{0.5cm}
\begin{tabular*}{\textwidth}{@{}l@{\extracolsep{\fill}}cccc}
\cline{2-5}
                 & \multicolumn{1}{c}{SM}
                 & \multicolumn{1}{c}{2H}
                 & \multicolumn{1}{c}{SUSY ($\not{\!\mbox{R}}$,R)}
                 & \multicolumn{1}{c}{Exotic quarks}\\
\cline{2-5}
B(t$\to$qZ)  & $\sim 10^{-13}$ & $\sim10^{-6}$ &
                 $\sim10^{-4},10^{-8}$ & $\sim 10^{-2}$  \\
\hline
\end{tabular*}
\vspace{0.5cm}
\end{table*}

\begin{eqnarray}
\label{lag4:vtq}
{\mathcal{L}}^{Vtq}_4&=& -g_s \bar{t} \gamma^\mu T^a t G_{\mu
a}-\frac{2}{3}e\bar{t}\gamma^\mu t A_\mu \\
& &-\frac{g}{\sqrt{2}}\sum_{q=d,s,b}\bar{t}\gamma^\mu
(v^W_{tq}-a^W_{tq}\gamma_5)q W^+_\mu + \mathrm{h.c.} \nonumber \\
& & -\frac{g}{2\cos \theta_W} \bar{t}  \gamma^\mu
(v^Z_{tt}-a^Z_{tt}\gamma_5)t Z_\mu \nonumber \\
& &-\frac{g}{2\cos\theta_W}\sum_{q=u,c}\bar{t}\gamma^\mu
(v^Z_{tq}-a^Z_{tq}\gamma_5)q Z_\mu + \mathrm{h.c.}. \nonumber
\end{eqnarray}
The first two terms are fixed by the unbroken gauge symmetry
$SU(3)_C\times U(1)_Q$. The charged currents are modified
$v^W_{tq},a^W_{tq}=\frac{V_{tq}}{2}
(1+{\mathcal{O}}(\frac{v^2}{\Lambda^2}))$, 
where $V_{tq}$ is the Cabibbo-Kobayashi-Maskawa (CKM) matrix
\cite{ckm}, 
receiving 
corrections of order $\frac{v^2}{\Lambda^2}$. The neutral currents are
similarly modified
$v^Z_{tt}=\frac{1}{2}-\frac{4}{3}\sin^2\theta_W+
{\mathcal{O}}(\frac{v^2}{\Lambda^2})
$, $a^Z_{tt}=\frac{1}{2}+{\mathcal{O}}(\frac{v^2}{\Lambda^2})$ and
$v^Z_ {tq},a^Z_{tq}={\mathcal{O}}(\frac{v^2}{\Lambda^2})$.   

These non-standard contributions come from the ${\mathcal{L}}_6$
operators in Table~\ref{operators:tab}. We follow the
 notation in Ref.~\cite{Buchmuller:1986jz} where 
$q$ stands for left-handed (LH) doublets and $u$
and $d$ for the right-handed (RH) singlets, $\phi$ is the scalar
doublet, $W_{\mu\nu}^I,B_{\mu\nu}$ are the usual $SU(2)_L$ and
$U(1)_Y$ 
gauge field strengths and $D_\mu$ is the covariant derivative. 
Other ${\mathcal{L}}_6$
operators involving the top quark do not contribute to
${\mathcal{L}}_4^{Vtq}$ at this order. As a matter of fact
${\mathcal{O}}_{d\phi}$
 does not contribute to this order either.
It redefines the down quark mass eigenstates, but this redefinition can
be reabsorbed in the CKM matrix. On
the other hand the operators ${\mathcal{O}}_{\phi W,\phi B,WB}$,
${\mathcal{O}}^{(1,3)}_\phi$ give flavour independent
corrections. They redefine the gauge couplings which are known to
fulfil the SM relations to an accuracy better than
$1\%$~\cite{Martinez:1999rs}. Hence, in
practice we are only interested in ${\mathcal{O}}^{(1,3)}_{\phi q}$,
${\mathcal{O}}_{u\phi,\phi u,\phi\phi}$. ${\mathcal{O}}_{u\phi}$
corrects the up quark mass matrix, although its diagonalization does
not modify the couplings in Eq.~\ref{lag4:vtq} at order~$
\frac{v^2}{\Lambda^2}$. Hence the corrections to the W and Z 
couplings come only from ${\mathcal{O}}^{(3)}_{\phi q}$ for charged
LH couplings; from ${\mathcal{O}}_{\phi \phi}$ for charged
RH couplings; from ${\mathcal{O}}^{(1,3)}_{\phi q}$ for neutral
LH couplings and from ${\mathcal{O}}_{\phi u}$ for neutral
RH couplings.

The size of the different corrections is dictated by the coefficients
of the relevant ${\mathcal{L}}_6$ operators, which depend on the
specific SM extension. Then one must wonder what new physics can
produce large top mixing. The order of magnitude of these coefficients
is fixed in order to reproduce, for instance, the branching ratios of
$t\to Z\;q$, $q=u,c$, in the full theory.
 Their small value for the two Higgs model and for
the supersymmetric models (see Table~\ref{br:models})
implies uninterestingly small ${\mathcal{L}}_6$ coefficients for these
two cases. On the contrary the allowed values in Table~\ref{br:models}
for simple SM extensions with exotic vector-like quarks make
important to discuss this possibility. Thus in the following section
we integrate out the simplest SM extension with an extra quark
isosinglet $T_{L,R}$ as an example.

It is worth before, however, 
to note that large top mixing can be induced by new
vector-like quarks and only by them. In Ref.~\cite{Arzt:1995gp} it was
shown that large ${\mathcal{L}}_6$ coefficients mean operators
generated by tree level integration of heavy modes 
and this implies for ${\mathcal{L}}_4^{Vtq}$ 
heavy fermions or heavy gauge bosons. As the mixing of the SM gauge
bosons is known to be typically small~\cite{Martinez:1999rs}, 
we are left with heavy vector-like quarks as the only
possibility~\cite{apvs:letters}.

\section{A simple example: one extra quark isosinglet $T_{L,R}$}

The simplest SM extension with large top mixing results from the
addition of only one charge $\frac{2}{3}$ quark isosinglet $T_{L,R}$
to the SM. In this renormalizable model the mixing of the new
vector-like quark with the SM quarks violates the GIM mechanism
introducing tree level Flavour Changing Neutral Currents (FCNC). This
extension is so simple that it is easier to work in the full theory in
this particular case, but it will be instructive to use the effective
Lagrangian formalism to guide the analysis of more complicated
cases. The complete integration of the general case will be given
elsewhere~\cite{apvs:letters}.

Let us assume without loss of generality canonical covariant
derivative terms, diagonal SM Yukawa couplings
$\lambda_i \bar{q}_i u_i \tilde{\phi}+\mbox{\textit{h.c.}}$, mixing
Yukawa terms $\lambda^\prime_i \bar{q}_i T_R
\tilde{\phi}
+\mbox{\textit{h.c.}}$ 
and a heavy mass term $M
\bar{T}_LT_R +\mbox{\textit{h.c.}}$, 
where $\lambda_i$ and $M$ are real and
$\lambda_i^\prime$ are in general complex.  
Then the integration of $T$ generates only new LH currents,
\textit{i.e.} only ${\mathcal{O}}^{(1,3)}_{\phi q}$ 
get non-zero
contributions. Now $\Lambda=M$. After SSB ${\mathcal{O}}^{(1,3)}_{\phi
q}$ give the ${\mathcal{L}}_4^{Vtq}$ corrections of order
$\frac{v^2}{M^2}$ through diagram (a) in Figure~1

\begin{equation}
{\mathcal{L}}_4^{Wtq}=-\frac{g}{\sqrt{2}} W^+_\mu A_{j k}
\bar{u}_L^j \gamma^\mu d_L^k+\mbox{\textit{h.c.}},
\end{equation}
\vspace{1cm}
\begin{eqnarray}
{\mathcal{L}}_4^{Ztq}&=&-\frac{g}{2\cos\theta_W}
Z_\mu(B_{jk} \bar{u}_L^j \gamma^\mu
u_L^k\nonumber \\ &&-\bar{d}_L^i\gamma^\mu d_L^i
- 2 \sin^2\theta_W J^\mu_{EM}),
\end{eqnarray}
where

\begin{equation}
A_{j k}=\left(\delta_{jl}+\frac{m'_jm^{\prime *}_l}{M^2}
\frac{m^2_j}{(-)^{\delta_{jl}}m_l^2-m_j^2}\right) V_{lk},
\end{equation}

\begin{equation}
B_{jk}=\delta_{jk}-\frac{m'_jm^{\prime *}_k}{M^2}, 
\end{equation}
and $m_i=v\frac{\lambda_i}{\sqrt{2}}$ and similarly
$m_i^\prime=v\frac{\lambda_i^\prime}{\sqrt{2}}$. The star stands for
complex conjugation.

\begin{figure}[ht]\label{fig:1}
\begin{center}
\begin{picture}(215,80)(0,0)
\DashLine(25,75)(50,50){5}
\DashLine(75,75)(50,50){5}
\Line(25,25)(50,50)
\Line(75,25)(50,50)
\Photon(50,50)(50,90){3}{3}
\put(19,75){+}
\put(73.5,75){+}
\put(43,0){(a)}
\DashLine(155,75)(180,50){5}
\DashLine(205,75)(180,50){5}
\Line(155,25)(180,50)
\Line(205,25)(180,50)
\DashLine(180,50)(180,90){5}
\put(149,75){+}
\put(203.5,75){+}
\put(176.25,91){$\times$}
\put(175,0){(b)}

\end{picture}\end{center}
\caption{${\mathcal{L}}_6$
 contributions after SSB to the dimension
4 effective Lagrangian ${\mathcal{L}}_4^{Vtq}$.}
\end{figure}
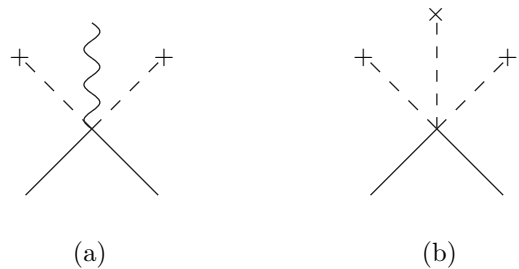

\begin{figure}
\begin{center}
\begin{picture}(200,80)(20,0)
\Line(45,40)(75,40)
\Line(45,42)(75,42)
\Line(25,30)(44,41)
\Line(95,30)(76,41)
\Photon(60,42)(60,80){3}{3}
\DashLine(45,40)(45,65){5}
\DashLine(75,40)(75,65){5}
\put(41.25,66){$\times$}
\put(71.25,66){$\times$}
\put(55,0){(a)}
\Line(155,40)(185,40)
\Line(155,42)(185,42)
\Line(135,30)(154,41)
\Line(205,30)(186,41)
\LongArrow(194,42)(204,36)
\DashLine(155,40)(155,65){5}
\DashLine(185,40)(185,65){5}
\put(151.5,66){$\times$}
\put(181.5,66){$\times$}
\put(165,0){(b)}
\end{picture}\end{center}\label{fig:2}
\caption{Quark isosinglet contributions to the dimension 4
 effective Lagrangians ${\mathcal{L}}_4^{Wtq}$, ${\mathcal{L}}_4^{Ztq}$.}
\end{figure}
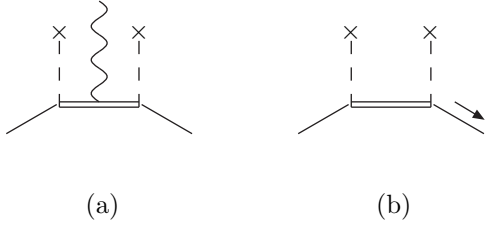

In the full theory the contributions from diagram (a) in Figure~2
correspond to a part of the contribution depicted by 
 diagram (a) in Figure~1. The other part
results from diagram (b) in Figure~2. This second diagram gives a
$\frac{v^2}{M^2}$ correction to $\bar{u}_{L}^i \not \! \partial
u_{L}^j$, where the arrow in the diagram stands for the
derivative. Then using the equations of motion in order to compare
with the ${\mathcal{L}}_6$ contributions in Figure~1, $\not \! \!
\partial u_{L}^j$ is replaced by a mass term, which will match with
the contribution of 
diagram (b) in Figure~1, and the remaining gauge terms in the
covariant derivative, which will complete the matching with diagram
(a) in Figure~1.

When confronted to experiment this simple model allows for a large top
mixing, in particular for instance $|B_{tc}|=\left|\frac{m_t^\prime
m_c^{\prime *}}{M^2}\right|\leq 0.082$ \cite{delAguila:1999tp}. 
This is almost one
order of
magnitude larger than the expected precision at
LHC, which corresponds to the branching
ratios in Table~3 \cite{Beneke:2000hk}.

\begin{table}[!ht]
\caption{Experimental limits expected at LHC for the top quark flavour
changing branching ratios.}
\vspace{0.25cm}
\begin{tabular}{ccc}
\cline{2-3}
& q=u & q=c \\ \hline
Br(t$\to$qZ)  & $10^{-4}$ & $10^{-4}$\\ \hline 
\end{tabular}\vspace{0.5cm}
\end{table}

\end{document}